# Silicon Detector Arrays with Absolute Quantum Efficiency over 50% in the Far Ultraviolet for Single Photon Counting Applications

Shouleh Nikzad[a], Michael E. Hoenk[a], Frank Greer[a], Todd Jones[a], Blake Jacquot[a], Steve Monacos[a], J. Blacksberg[a]
Erika Hamden[b], David Schiminovich[b],
Chris Martin[c] and Patrick Morrissey[c]

[a]Jet Propulsion laboratory, California Institute of Technology, Pasadena, California, 91109
[b]Columbia University, New York, NY 10025
[c]California Institute of Technology, Pasadena, CA 91125

## ABSTRACT

We have used Molecular Beam Epitaxy (MBE)-based delta doping technology to demonstrate near 100% internal quantum efficiency (QE) on silicon electron-multiplied Charge Coupled Devices (EMCCDs) for single photon counting detection applications. Furthermore, we have used precision techniques for depositing antireflection (AR) coatings by employing Atomic Layer Deposition (ALD) and demonstrated over 50% external QE in the far and near-ultraviolet in megapixel arrays. We have demonstrated that other device parameters such as dark current are unchanged after these processes. In this paper, we report on these results and briefly discuss the techniques and processes employed.

+ Electronic mail: Shouleh.Nikzad@jpl.nasa.gov

Ultraviolet, Optical, and near infrared imaging has wide reaching applications in space, medical, commercial and communication fields. High efficiency ultraviolet photon detection is essential for many instrument and sensor concepts and has been the focus of recent research endeavors in astrophysics, planetary science, biology, and biomedical fields. For example, future NASA UV/Optical missions beyond Hubble Space Telescope (HST) and the Galaxy Evolution Explorer (GALEX) will require significant detector advances, particularly in quantum efficiency, noise, resolution, and number of pixels in order to result in major new scientific impacts.

The current UV detection technologies can be classified into two major categories: 1) Solid-state devices based on silicon or wide bandgap semiconductors and 2) a detector that is a combination of a photo-emissive device, (i.e., a photocathode), a gain component, and an electron detector. Electron bombarded CCDs (EBCCDs) and microchannel plates (MCPs) are in the latter category. Traditionally, for space and laboratory applications, researchers have chosen MCPs because of their low noise and high gain which enhances the signal to noise ratio at low light levels and enables photon counting. This is despite their relatively low UV quantum efficiency QE (~10%) and the requirement to utilize high voltage power supplies for their operation. Solid-state detectors, on the other hand, offer significant advantages in size, mass, power, reliability and manufacturability compared to the vacuum tube based technology of MCPs. Therefore, using solid-state detector arrays improves instrument compactness and reduces instrument complexity, both of which are at a premium especially in instruments designed and built for space and biomedical applications. However, to surpass the performance of MCPs and enable major new scientific impacts, what is required is a solid-state UV detector with stable, high QE, low noise, and moderate gain.

Wide bandgap materials such as gallium nitride (GaN) are intrinsically insensitive to visible photons and are being formed into UV detector arrays as well as UV photocathodes. To form UV detector arrays, a multilayer structure of GaN is grown by a technique such as

MOCVD or MBE on a sapphire substrate. An array of p-i-n diodes is then fabricated from this material and hybridized to a CMOS readout array. There has also been report of success with wide band gap avalanche photodiode arrays possessing gain.[1-3] On the other hand, an immense investment has been made in order to produce silicon detectors with very low noise, low dark current, and very large imager formats. In addition, new techniques such as Electron Multiplied Charge-Coupled Devices (EMCCDs)[4-6] with lateral gain and low noise CMOS sensors with in-pixel avalanche photodiodes are being developed[7, 8] so that silicon sensors continue to improve and have now become viable for photon counting applications.

Using silicon arrays for UV detection historically has posed a challenge. Because of the shallow absorption length of UV photons into the material, silicon imaging arrays are not suitable in a conventional front-illuminated configuration for UV detection. Instead silicon arrays must be operated in a thinned, back-illuminated configuration, in which photons are incident on the bare silicon surface opposite from the circuitry. While back illumination avoids absorption and scattering in the gate and pixel structures on the front surface, back-illuminated silicon arrays require effective methods of surface passivation to achieve high, stable quantum efficiency especially in the ultraviolet. Without effective surface passivation, traps at the $Si-SiO_2$ interface can dynamically interact with photo-generated charge, resulting in poor quantum efficiency, unstable response (quantum efficiency hysteresis), and high dark current. Various surface passivation methods have been developed for back-illuminated detectors in order to improve the UV response, reduce surface-generated dark current, and stabilize the overall quantum efficiency. Ion implantation techniques and chemisorption techniques applied to thinned backside-illuminated charge-coupled devices (CCDs) have achieved high quantum efficiencies in the near ultraviolet region of the spectrum (above 250 nm).

Delta doping technology, developed at NASA's Jet Propulsion Laboratory (JPL), California Institute of Technology, using molecular beam epitaxy (MBE) uniquely achieves atomic-scale control over the surface bandstructure of a silicon imaging array, resulting in near-100% internal quantum efficiency from the EUV through near infrared regions of the spectrum, very low surface-generated dark current, and elimination of quantum efficiency hysteresis[9-11].

In this paper, we report the demonstration of a high QE, far ultraviolet solid-state detector array with low noise and moderate gain, by combining three independent technologies in a new way. EMCCDs provide the necessary gain and low noise while delta doping and ALD provide atomic scale control over interface properties to achieve high and stable QE. We have achieved near 100% internal QE or reflection limited response using MBE delta doping of EMCCDs. Furthermore, we have enhanced the QE in the deep ultraviolet by depositing single-layer antireflection (AR) coatings using precision control of atomic layer deposition (ALD). In certain cases, we have also employed thermal evaporation for AR coating deposition. Features of this deep UV detector technology address the key requirements for UV photon counting detection with solid-state detector arrays. Single photon counting is achieved by the electron-multiplied design and efficient detection of UV photons is achieved by using an ultrathin p-type delta-doped layer to passivate and accumulate the back surface. Because the delta doped surface is terminated by silicon and its native oxide, further enhancement of QE using antireflection coatings is possible without significant constraints based on the starting surface. The challenges of developing AR coatings for this part of the spectrum (*i.e.*, 100-300 nm region) include a rapidly changing index of refraction, material limitations for having both low absorption and high index, constraints on deposition methods that at once offer high quality ultrathin layers without affecting the delta doped layer or the front-side collection and readout circuitry of the device.[12, 13] We report more than 50% measured external quantum efficiency (QE) in the 100-300 nm spectral range by designing and depositing AR coatings using ALD and thermal evaporation.

Electron-multiplied CCDs have been invented at e2v in their Low Light Level L3Vision CCD design.[5] This approach allows leveraging all the advantages of the mature CCD technology while enabling single photon detection by adding a gain register at the end of the charge transfer prior to the readout amplifier. At each stage of the final gain serial shift register, higher gate voltages in the second serial clock phase cause avalanche effect that produces a small gain with a final gain of more than one thousand by the end of the register.

Commercial EMCCDs, i.e., 1024 x 512, 16-micron pixel L3CCDs (e2v CCD97s) with 20-micron epilayers were used in this effort. For our purpose, we thinned to 8 $\mu$m by removing the p+ substrate and approximately 12 $\mu$m of the epilayer using a multistep process starting with chemical mechanical polishing (CMP) process followed by chemical thinning to achieve a smooth and highly-specular silicon surface. Following the thinning, a series of solvent cleaning steps were used to eliminate residual organic material used in processing steps such as photoresist and waxes. The devices were then delta doped following the processes described in detail elsewhere.[9, 14] Briefly, a 2.5 nm epitaxial delta layer of highly boron-doped silicon is grown on the back surface of the device using MBE. After delta doping, the devices were used for a variety of antireflection coating processing. To establish the success of combining the delta doping process with electron multiplied CCDs, the first devices were mounted in packages and wire bonded without any further processing. Devices were characterized in a vacuum UV system equipped with two sources of deuterium and tungsten allowing characterization from 120-700 nm. The system was fully calibrated including a NIST calibrated diode to accurately measure the flux of the photons that reach the devices. Details of this characterization system are described elsewhere.[15, 16]

For the further enhancement of the quantum efficiency, ALD was used for the deposition of ultrathin layers of hafnium oxide and aluminum oxide. The layers were modeled using TFCalc™ and they were deposited on the back surface of thinned delta doped conventional 1kx1k CCDs. These devices provided a suitable platform for the demonstration and optimization of the AR coatings. The magnesium fluoride AR coating was deposited using thermal evaporation. Work is now underway to deposit the fluoride film by ALD as well.

We first present the result of a bare delta doped L3CCD. It is important to do so because first, it serves as an experimental control for AR coatings and second it validates the surface passivation. A back illuminated device should respond at the reflection limit of silicon if there are no other losses for the photo-carriers. If the backside potential well is not eliminated (*i.e.*, if the surface states are not fully passivated), quantum efficiency hysteresis (QEH) will occur. QEH is highly undesirable, especially in space imaging and spectroscopy applications. For example, the Hubble Space Telescope Wide Field Camera requires a periodic, intense flood illumination to remove this QEH. This has been discussed in detail elsewhere.[17, 18] Figure 1 shows the quantum efficiency of a delta doped electron multiplied CCD or delta doped L3CCD. For comparison, silicon transmittance, or 1-R where R is reflectivity of a bare silicon surface, also has been plotted [Hamden, and references therein].[12] It is important to note that in reporting our data, we account and correct for the multiple electron-hole pair, i.e., quantum yield (QY) produced by higher energy UV photons and present true QE or the detected photons as a percentage of incident photons. At higher wavelengths or lower photon energies, the QY is approximately unity and at wavelengths shorter than 360 nm[19], this number is qualitatively proportional to the incident photon energies. This proportionality has been modeled and measured for higher photon energies including x-rays and has been reported for silicon by different group as a constant number or as a function of incident photon energy.[15, 19-21] A detailed discussion can be found elsewhere.[16, 20, 22] Accounting for QY and normalizing to that number is important because failure to do so results in erroneously high QE (*e.g.*, it could be over 100%). In figure 1, we directly demonstrate this phenomenon by presenting our data in three ways: 1) The ratio of measured electrons to the number of incident photons (analyzed signal not normalized to QY), 2 & 3) The ratio of detected photons to the number of incident photons using two different QY set of values to correct and normalize for multiple electrons per photon in the measured signal. As shown in the figure, with either correction and within the uncertainties of the measurement, the device is responding at reflection limit or nearly 100% internal quantum efficiency. As shown in the figure, the detected not normalized for QY will result in a false high QE in error.

The response of another delta doped L3CCD array was measured after the device was partially AR coated by shadow masking using a silicon piece to cover a portion of the device. An $Al_2O_3$ layer was modeled to cover 170-200 nm region and allow enhancement of the response so that the final measured quantum efficiency is >50% over a significant wavelength range. Figure 2 shows the QE response of the bare delta doped region and the AR coated region of the same device. As shown in the figure significant enhancement is achieved due to the single $Al_2O_3$ layer film.

Other antireflection coatings were designed to cover the rest of the 100-300 nm region of the spectrum and to conserve resources these were all demonstrated on thinned and delta doped conventional CCDs with Cassini mission heritage which are readily available to our group at JPL. In figure 3 the detectors responses are plotted as a function of incident photon wavelength showing that QE > 50% is achieved in the ultraviolet. The response of an AR-coated and delta doped L3CCD is also plotted and within the uncertainties is in agreement with the other device measurement. For comparison, the response of the MCP detectors flown on GALEX is also plotted.[23, 24]

In addition to quantum efficiency measurements we characterized the device for dark current, another measure of a well passivated back illuminated silicon array by combining three technologies. Figure 4 shows the results of dark current measurement as a function of temperature for both regions of a single device that was partially AR coated in one region. Also, plotted on the figure is e2v model of device dark current as a function of temperature presented as a dashed line.[25] It is shown in the figure that the device behaves according to the model and both the AR coated and bare regions delta doped array follow the expected behavior as a function temperature.

In conclusion, we have demonstrated high quantum efficiency in the far ultraviolet in solid-state arrays that possess gain and low noise. We have achieved this by delta doping and AR coating of an electron multiplied CCD. The bare delta doped array resulted in near 100% internal QE. We have further enhanced the QE by developing advanced far UV AR coating layers by using single layers in different portions of the spectrum. The significance of these results is enabling solid-state detectors for achieving high quantum efficiency in far ultraviolet in a platform of a low noise device that possesses gain and therefore enables single photon counting. Additionally, unlike photo-emissive devices such as MCPs or EBCCDs, the combination of these techniques provides a powerful detector in the ultraviolet and far ultraviolet without use of high voltage.

**Acknowledgements**

The work presented in this paper was performed by Jet Propulsion Laboratory (JPL), California Institute of Technology under a contract with NASA. We gratefully acknowledge the generous collaborative effort by e2v Inc. and helpful discussions with Peter Pool and Paul Jerram of e2v.

# Figures

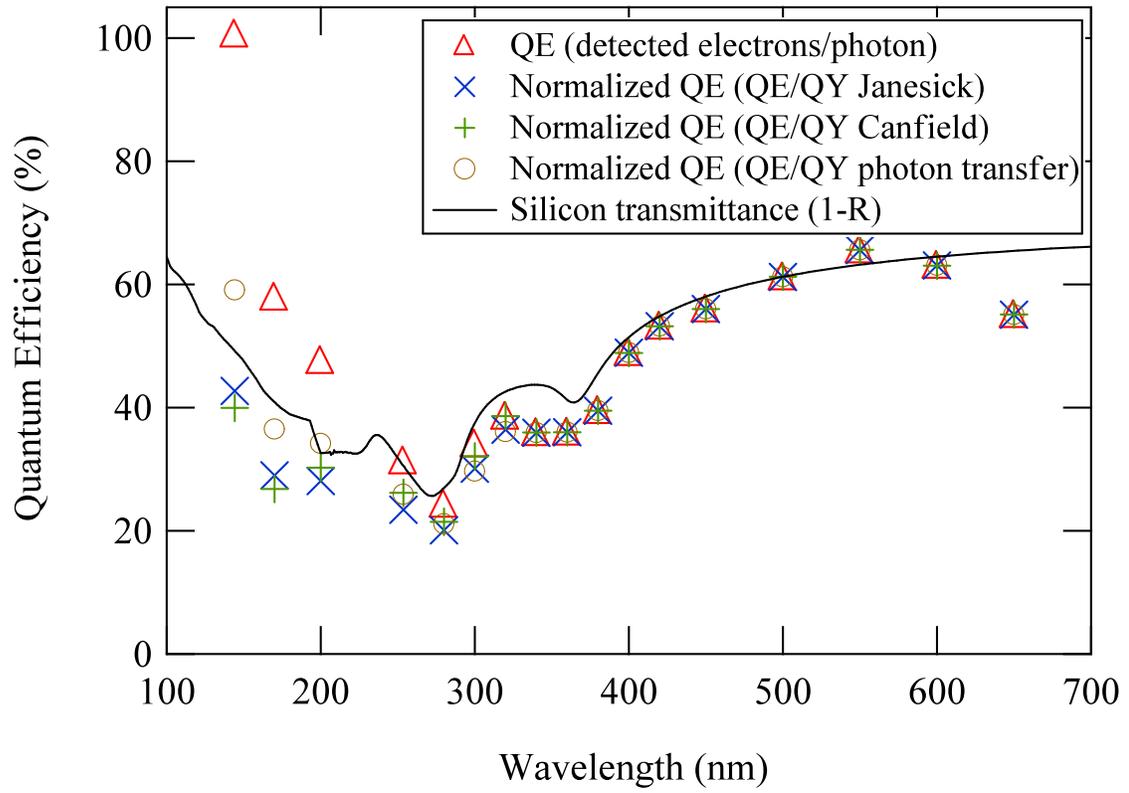

Figure 1

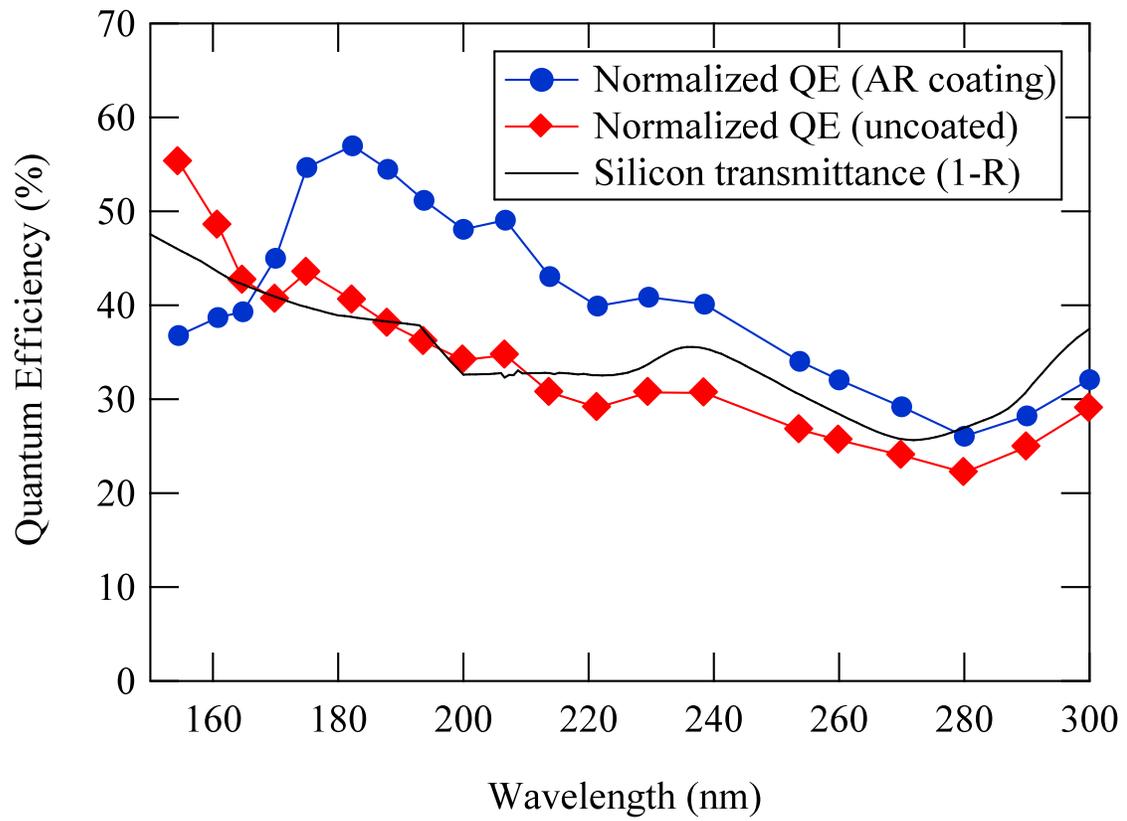

Figure2

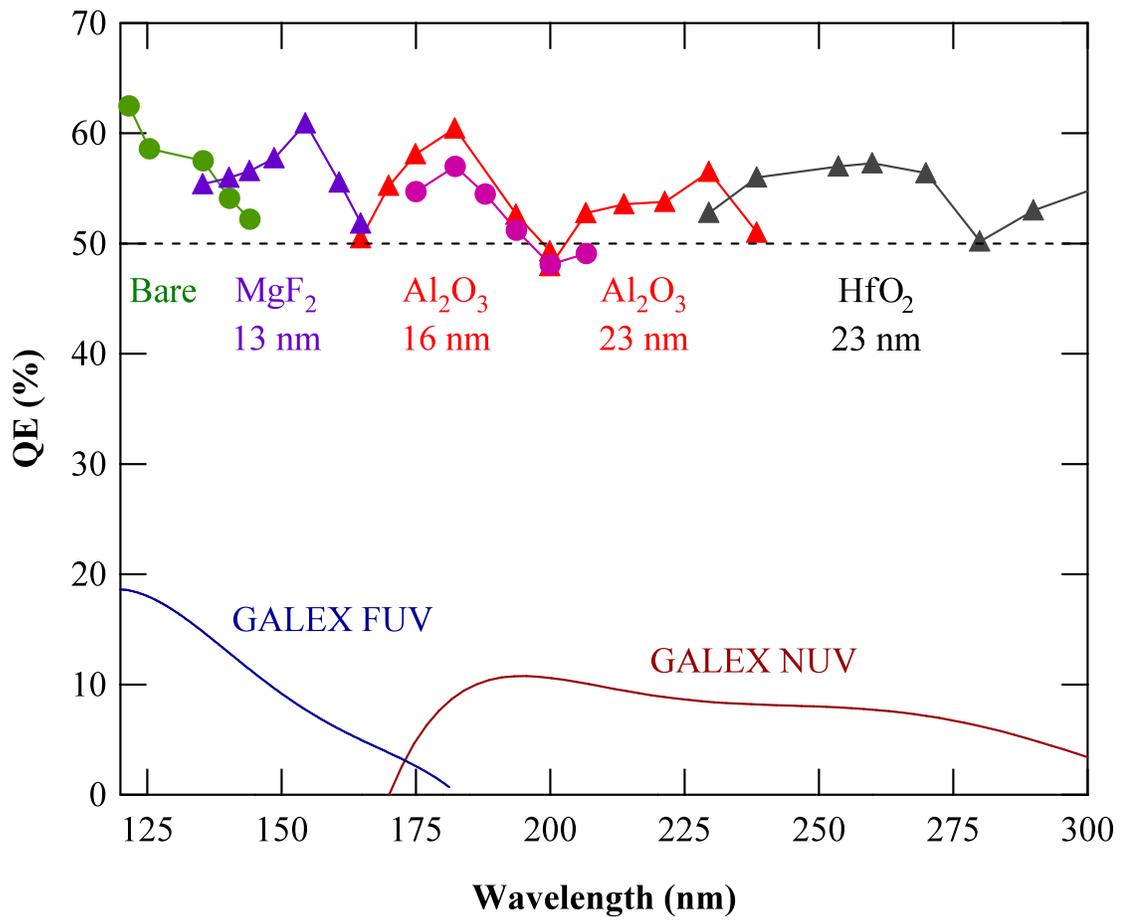

Figure 3

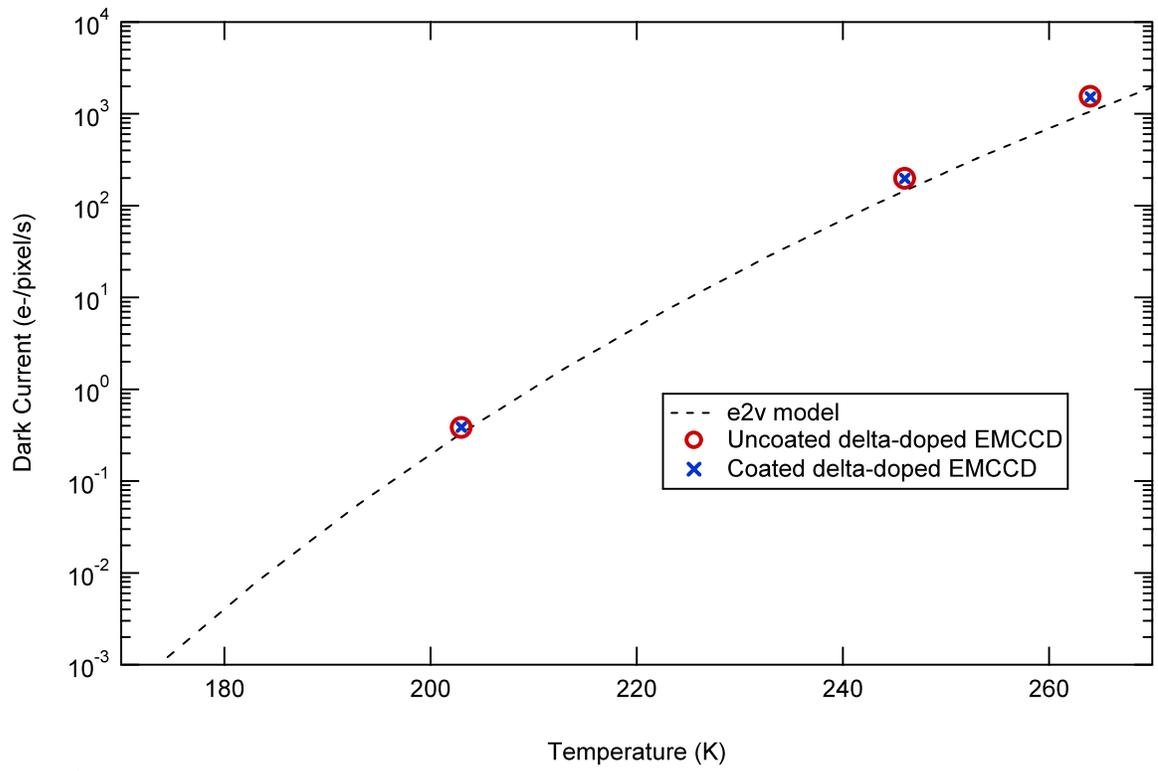

Figure 4

**Figures Captions**

Figure 1. Quantum efficiency of a 512 x 1024 delta-doped electron multiplied CCD. The CCD is an e2v L3CCD that was thinned and delta doped at JPL. The solid line is the theoretical reflection limit of silicon surface. Data is corrected for multiple electron-hole pair production using two different approximations. Either correction gives a reasonable agreement with transmittance (1-Reflection) of silicon.[12] A conservative estimate in the uncertainty of the measurements is 10%.

Figure 2. Far ultraviolet and ultraviolet response of a delta-doped electron multiplied L3CCD. The response of the coated region shows significant enhancement in efficiency over the uncoated region in the region targeted by the aluminum oxide AR coating. The uncoated region QE is in good agreement with the silicon transmittance data within the uncertainties of measurements.[12,16] The QY method used to determine the true QE, (i.e., fraction of photons detected) was direct measurement at each wavelength in our laboratory using the photon transfer technique.[15, 21]

Figure 3. Response of the AR-coated, delta doped arrays in the entire 100-300 nm region. Each region was coated by one simple coating. Further enhancements can be achieved by adding additional complexity in the coatings. In the third band, the response of AR coated, delta doped EMCCD (16nm $Al_2O_3$) is plotted on the same plot as the conventional CCDs.

Figure 4. Dark current measurements for a partially AR coated delta-doped electron multiplied L3CCD. The coated and uncoated data are derived from measurements of different regions of the same device, with the uncoated region defined by a contact shadow mask during the AR

coating deposition process. Both sets of data follow the dark current model developed by e2V as a function of temperature for their devices.[25]